# Tale of Tilted Washboards of Potential Energy

## Celso L. Ladera and E. Stella

## Departamento de Física, Universidad Simón Bolívar

## Caracas 1089, Venezuela

**Abstract**. Tilted washboard potentials are periodic functions of potential energy whose graphs remind the profiles of inclined washboards for scrubbing clothes. Practically unknown until the 1970´s, washboard potentials are at present ubiquitous seminal landscapes across diverse physics researches and technologies now receiving significant attention. However, washboard potentials are still rarely presented to science and engineering undergraduates; perhaps they could hear about those when first studying Josephson junctions. Aiming to promote the presentation of such potentials to undergraduates, we here consider cases of washboard potentials in: *classical mechanics*, *quantum physics* and *optics*. The quantum case is the superconducting Josephson junction whose *phase* quasi-particle has a tilted washboard potential leading to important applications, *e.g.* the *phase qubit* in quantum computation. Optical cases are *electro-optical* potentials generated with laser beams in which thousands of atom electric dipoles may become confined or evolve, potentials being generated either with laser Bessel beams, or with laser standing waves, whose periodic potential profile is implicitly tilted or may be experimentally adjusted to become tilted. Electro-optical potentials are presently used in fast-paced fundamental research, *e.g.* Brownian particle motors, atom trapping in optical lattices, optical atomic clocks, protein transport in biological cells, and in condensed matter physics to name a few.

## 1 INTRODUCTION

Paying attention to the potential energy function of classical physics systems is a must for understanding and analytically representing the motion and physics interactions of such systems. This is illustrated by well-known cases *e.g.* the parabolic potential of the simple harmonic oscillators, and the gravitational potential of a planet. Further examples of relevant energy potentials are considered in modern physics and quantum mechanics to explain important quantum interactions, *e.g.* the Lennard-Jones energy potential between the atoms of diatomic molecules [1] and the Square-well potentials and the Dirac delta potential [2,3]. In this work we consider the lesser known *tilted washboard energy potentials* [4-8] of some physics systems, represented by functions *U(x)* of the form:

$$U(x) = -A\,x - B \cos x, \quad (1)$$

where *x* is the pertinent coordinate of the system, *A* and *B* being real numbers, in the proper physics units. Such kind of potentials began to be presented in textbooks by the 1970´s years, the first example perhaps being presented by M. Tinkham in his book on superconductivity [4]. The graphs of these potentials are tilted cosine functions, which tilts are determined by the first constant $A$ in Eq (1). Two examples are graphed in Fig. 1, the second one showing a larger tilt. Notably, tilted washboard potentials show multiple local *wells* of energy, which depths happen to depend on the ratio $B/A$ of the coefficients in Eq (1). As shall be seen, in this work, such wells are the key for important quantum physics technologies based on washboard potentials.

$$\textbf{(a)} \quad U_1(x) = -0.1\,x - \cos x; \text{ and} \qquad \textbf{(b)} \quad U_2(x) = -0.4\,x - \cos x, \qquad (2)$$

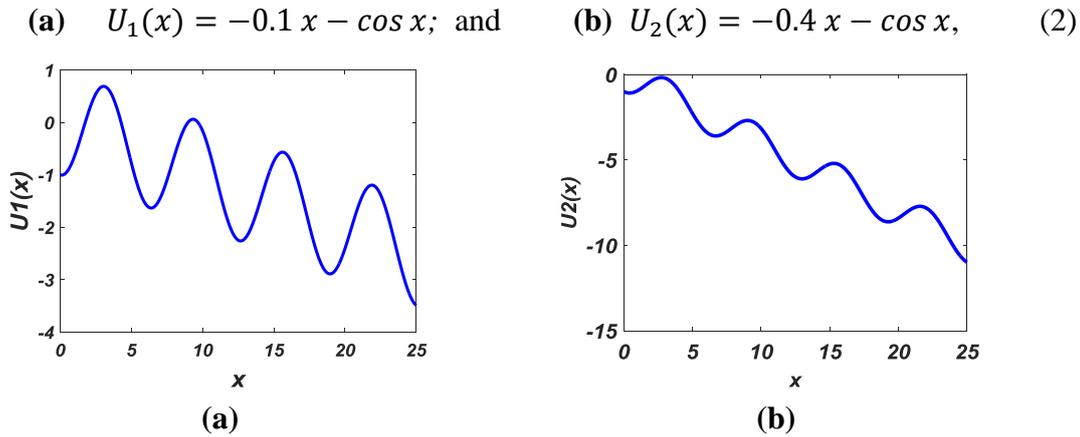

(a)                        (b)

**Fig.1** Examples of tilted washboard potential functions: (a) $U_1(x) = -0.1\,x - \cos x$ ; (b) $U_2(x) = -0.4\,x - \cos x$; the second of larger tilt but with wells of lesser depth.

If $B=A$ in Eq (1) the minimae of the local wells of the tilted washboard potential become just horizontal *inflection points*: equilibrium points disappearing and the depth of the local potential wells becoming zero [4, Ch. 6], a fact that implies a particular motion of the physics system, or particle, whose potential energy is represented, *e.g.* if a bead it will then evolve down to reach the bottom of its potential energy. In the sections below we shall present and study examples of important physics systems whose potential energy has the profile of inclined washboard functions *e.g.* the superconductive Josephson junctions [3-9] in Section 3.

This work pursues two goals: The first being educational: (i) To offer up-to-date motivating physics examples of these tilted potential energy functions, and how to treat them analytically; (ii) Secondly, at the cores of the mathematical-physics models and experimental work, of some present fast-paced physics research and key technologies, one finds such tilted washboard energy potentials. For instance, the *phase qubits* of quantum computers, the planar arrays of Josephson junctions, energy potential for studying ultra-cold gases of atoms, and *optical lattices*. As already recognized [10], washboard like potentials are *ubiquitous* and *seminal landscapes* across some present research fields and applications, including the extremely accurate optical atomic clocks, far more accurate than the caesium atom clock standard. Thus, modesty ahead, the second aspiration of this work is to offer undergraduate's students and instructors a few beach-bridges to approach these present research and relevant technologies, that they might eventually consider as subjects of their eventual Bachelor or Master theses (examples of such theses do exist).

Thus, we begin Section 2 illustrating with two simple classical mechanics examples: **(i)** the well-know driven and damped pendulum coupled to a pulley; **(ii)** a particle moving along a line whose potential energy is represented by a known tilted washboard function. In the first example the 2$^{nd}$ order differential motion equation of the pendulum will be first written, and from it we shall then derive the pertinent tilted washboard potential function of the pendulum.  In the second example −to emphasize the importance of the inclined washboard potential concept − we shall analyze the motion of a particle in the opposite sense:  Assuming its washboard potential function to be known, to then derive the motion differential equation of the particle.

Section 3 shall be devoted to the tilted washboard potential of an interesting and useful quantum system: the superconducting *Josephson junction* [3-9] proposed by B. D. Josephson in his PhD thesis [11]. As originally conceived the physics model of that junction (Fig. 2) is a tiny  device (say area ~500 μm$^2$) built with *two metallic electrodes*, A and B, separated by a *very thin layer of an insulating material* (a few nm wide)  *e.g.* made with aluminium electrodes and a very thin layer of its oxide: *Al – Al$_2$O$_3$ –Al*.

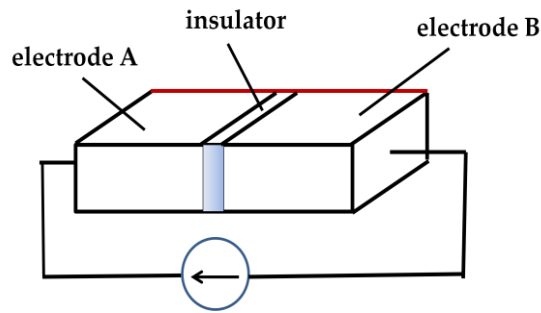

**Fig. 2** Scheme of a superconducting Josephson junction: A and  B are two metallic super conducting electrodes separated by a thin layer of an insulator.

Superconductivity and the Josephson junction are subjects currently presented in undergraduate modern physics courses and textbooks [3, 8]: When cooled down to liquid Helium temperatures (~4 K) and below, the two metallic electrodes of a *real* junction  become  superconductors *i.e.* their *electrical resistance reaching  zero value*, and as Josephson predicted: *Without any voltage applied*  between the superconducting electrodes, the electrical charge carriers in the electrodes should cross the thin insulating layer of his junction as a supercurrent  $I_J$. This is known as the *Josephson Effect* and has been experimentally confirmed. It is known [4,5,7,9]  and presented in modern physics textbooks that the currents in superconductors are not of single electrons, instead they are *pairs of electrons* of total charge *q=2e*, known as *Cooper pairs*.

In his thesis Josephson also proposed that: when cooled at superconducting temperatures, the *supercurrent $I_J$* of Cooper pairs of electrons that *tunnel* through the insulating layer of a junction, with no voltage applied, is given by the equation:

$$I_J = I_c \sin \varphi \qquad (3)$$

where $I_c$ is called the *critical current*, a parameter value typical of each junction, and $\varphi$ is the *phase difference* between the quantum wave-functions [1-3] of the Cooper pairs at the two superconducting junction electrodes. In fact, this constant $I_c$ is the *maximum super-current that the junction may support*. The phase $\varphi$ is actually the most important variable of the Josephson junction mathematical-physics model, and in Section 3 we shall demonstrate that a *real* Josephson junction has a potential energy whose profile is precisely a tilted washboard potential function of the junction *phase $\varphi$* (Fig 4), which is then envisaged as if it were *the position of a phase quasi-particle*.

In his model of the junction Josephson also predicted that if a voltage difference $V(t)$ is applied across a junction then the change-rate $d\varphi/dt$ of the junction phase should be proportional to that voltage:

$$V(t) = \left(\frac{\hbar}{2e}\right)\frac{d\varphi}{dt} , \qquad (4)$$

the proportionality constant being $\frac{\hbar}{2e}$, where $\hbar$ is Planck´s constant divided by $2\pi$. Also note that Eq (4) may be rewritten as $V = \frac{h}{2e} f_J$, in terms of the so-called *Josephson frequency constant $f_J$* of the junction super-current:

$$f_J = \frac{2e}{h} V \qquad (5)$$

Josephson junctions have become an important asset for a number of present technologies that are based on superconductivity and quantum mechanics *e,g.* in quantum computers, superconductive quantum interference devices (SQUID [2]), ultra-sensitive magnetic field detectors, and even the implementation of a *Volt standard* [12]. In Section 3 we shall present a model of the Josephson junction *Phase qubit*, the simplest implementation of a *qubit* used in quantum computers, and whose operation is based precisely on the potential wells of the tilted washboard potential of the Josephson junction (see Fig. 4 below) and its phase quasi-particle $\varphi$.

A typical DC voltage $V$ applied to a junction is $\sim 100\mu V$, and since the Josepson proportionality constant in Eq. (5) is $\frac{2e}{h} = 483593.4 \frac{MHz}{V}$, it implies that microwaves of frequency $f_J \sim 50 GHz$ are generated by the junction for that applied voltage. Conversely, when irradiated with $\sim 50 GHz$ microwaves the junction shall exhibit a voltage value $\sim 100\mu V$. Since microwave frequencies can be measured with extreme accuracy then Josephson junctions can perform as *frequency-to-voltage* converter, and can be used to implement a *Volt Standard* in national bureaus of standards, *e.g.* at the BSI (U.K), NSI (USA), NPL (Geramny)

Section 4 is devoted to the case of tiny dielectric particles, *e.g.* silica beads, interacting with the electric field of a laser light beam. Such interaction occurs because the field induces an *electric dipole* [13] inside the particle. As well-known, opposite equal charges *(+q,-q)* separated at a distance *d* within a dielectric (insulator) material constitutes an *electric dipole*. A far more complex electric dipole is induced inside a dielectric particle by an electric field. The interaction of such induced electric dipoles with the electric field that induced it, is already presented in textbooks *e.g.* [9, Vol 2] and [14]. Applications of the *optical* forces on dielectric particles dipoles by electric fields of *light beams* are subject of increasing interest since the pioneering work of A. Ashkin in the 1970 on the so-called *optical tweezers* [15].

In effect, if a dielectric particle is placed in a region of high intensity in a laser beam of *wavelength* $\lambda$, then the laser electric field $E(\omega, r)$ will induce a partial separation of charges inside the particle: It will induce an oscillating electric *dipole moment* vector $\boldsymbol{d}(\omega, \boldsymbol{r}) = \alpha(\omega)\, \boldsymbol{E}(\omega, \boldsymbol{r})$, where $\alpha(\omega)$ represents the *complex* particle *polarizability* that *oscillates with the field frequency* $\omega$. The *polarizability* represents, or "tells", us how the electrical charges get distributed inside the interior of the dielectric particle and how they oscillate. This distribution of charge inside the dielectric particle create a *vector dipole* $\boldsymbol{d}(\boldsymbol{r})$ in it [9 (Vol.2, ch.6); 10, 14]. Since $\omega = 2\pi(c/\lambda)$ the induced *polarizability* $\alpha$ *depends on the* laser wavelength $\lambda$, a dependence now being exploited in advanced technologies, *e.g. optical atomic clocks* (Subsection 4.2). The particle induced dipole $\boldsymbol{d}(\boldsymbol{r})$ shall further interacts with the laser electric field $\boldsymbol{E}(\lambda, \boldsymbol{r})$: the dielectric particle then acquiring an *electro-optical potential energy* given by:

$$U(r, \omega) = \alpha(\omega)\, |\mathbf{E}(\boldsymbol{r})|^2 \qquad (6)$$

where $|\mathbf{E}(\boldsymbol{r})|^2 = I_l$ is the laser irradiance, or intensity (in W/m$^2$). The particle ends up being confined in this electro-optical potential *U(r)* by the electric force given by $\boldsymbol{F} = -\nabla U$, a confinement analogous −of course not equal − to the confinement of a satellite in the gravitational potential energy of a planet.

The irradiance of the laser beam used in these induced electric dipole interactions is *spatially periodic*, so that the electro-optical potential *U(r)* that is created is also periodic, and may resemble a washboard potential. In important cases of application this periodic potential results already tilted (see sub-Section 4.1): its profile being similar to a tilted washboard energy potential, where dielectric particle may result confined in the potential wells. As shall be explained below in Section 4, that interaction is also the working principle of the micro-manipulation of tiny particles in the two types of electro-optical washboard potentials mentioned in Section 4.

Most laser sources, used in undergraduate optics experiments and lecture demonstrations, *e.g.* He-Ne and diode lasers, emit coherent light beams whose transverse light irradiance is given by a square *Gaussian exponential* function $I(r)=I_0\, exp\,[-2r^2/w^2]$, *w* being the beam *waist* width (Fig.3a). The interactions of particle dipoles with the electric field of light beams (sub-Section 4.1) demand laser beams whose transverse irradiances should instead be proportional to the square of a *Bessel´s* *function* [16] (Fig.3b). These laser beams are called therefore *Bessel´s beams* [17] and up-to-day they are not usually presented in undergraduate optics courses and textbooks.

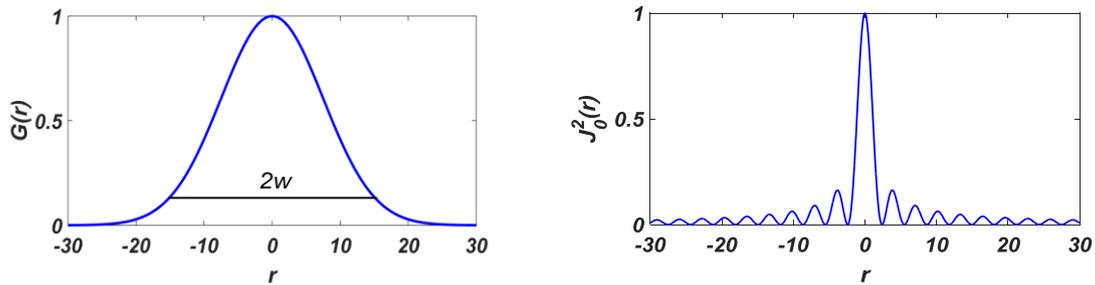

**Fig 3** Transverse *normalized* radial irradiances *vs* radius *r*: (a) laser Gaussian beam *G* and waist *w*; (b) a $J_0$ *Bessel* laser beam

Thus, a zeroth-order *Bessel beam* is a beam of light whose *transverse intensity profile* is quasi-periodic and represented by the square of *Bessel´s* $J_0$ function (Fig. 3(b)). We shall explain how Bessel´s beams can be used to generate challenging cases of washboard potentials where tiny particles are transported, or can be trapped, leading to frontier research and technological applications with particles such as atoms, and even living cells. As shall be clearly explained in Sub-section 4.1 the *Bessel-function* washboard potentials profiles are obtained implicitly tilted since they are actually generated in the laboratory using a Gaussian laser beam, or can be purposefully tilted by experimentalists with simple and minor adjustments of their optical set-ups.

Likely, readers have seen the appealing experiments titled *Acoustic Levitation* shown in Internet videos, in which small Styrofoam spheres appear *levitating* along a vertical column of air, actually along *sound generated standing waves*: the spheres being *trapped in the periodic low pressure wells* of the sound standing waves. Interestingly, as explained in Sub-section 4.2, analogous experiments for trapping sub-microscopic particles with *light standing waves* are also easy to set up. Light standing waves are a second case of electro-optical particle confinement, generated this time from the interaction of induced electric dipoles with the light standing waves generated with two counter-propagating laser beams [18]. The two beams producing spatially periodic arrays of electro-optical potential wells of equal depth (Fig 14).

In the laser light standing waves research and applications, the dielectric particles are sub-mcroscopic size electric dipoles that end-up being confined in the narrow standing wave wells, like *eggs in a carton*. This arrangement of trapping wells is called an *optical lattice*, and are periodic electro-optical potentials that again can be easily tilted by the experimentalists to make them to perform as tilted washboard potentials. Optical lattices are now at the very front of physics research *e.g.* the key element of very accurate atomic clocks is an optical lattice (accuracy of a second lost or gain in $10^9$ or moe years; see Sub-section 4.2) that can be even portable.

## 2  Two examples of mechanics tilted washboard potentials

In introductory physics, after the simple harmonic oscillator with its well known *parabolic* potential, the planar damped and driven pendulum oscillator attached to a pulley is usually presented (Fig. 4(a)), its motion differential equation (Eq.7) derived [5, 8,19] by applying Newtonian mechanics. In effect, if *m* is the mass of the pendulum bob, *L* its length, $\theta$ its angular position *w.r.t.* the vertical, then its motion equation is:

$$\tau = Y\ddot{\theta} + \eta\dot{\theta} + mgL\sin\theta \qquad (7)$$

where $Y=ml^2$ represents the bob moment of inertia, $\tau = MgR$ the torque on the pulley of radius *R*, *M* the mass hanging from the rope, and $\eta\dot{\theta}$ the viscous or damping term:

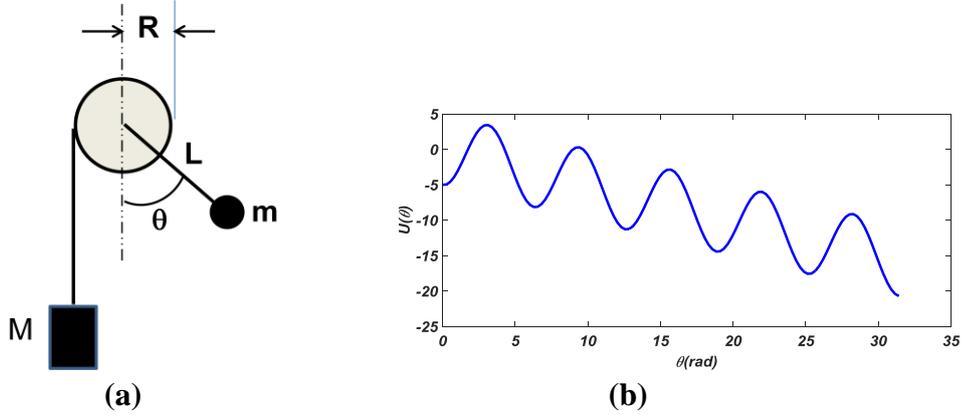

(a)                                                                 (b)

**Fig. 4 (a)** A damped driven pendulum of mass *m* attached to a pulley of radius *R*  **(b)** An example of its tilted washboard potential (after Eq (11)): $U(\theta) = -0.5\,\theta - 5\cos\theta$

To derive the potential energy function of this oscillator we begin simply rearranging Eq. (7) and multiplying it by the angular velocity $\dot{\theta}$ to get:

$$Y\ddot{\theta}\,\dot{\theta} + \left[-\tau\dot{\theta} + \dot{\theta}(mgL\sin\theta)\right] = -\eta\dot{\theta}^2 \qquad (8)$$

Which may be rewriten as

$$\frac{d}{dt}\left[\frac{1}{2}Y\dot{\theta}^2 + (-\tau\theta - mgL\cos\theta)\right] = -\eta\dot{\theta}^2 \qquad (9)$$

Note that the square bracket in the *l.h.s.* of Eq. (9) is the sum of the kinetic energy of the bob, plus its potential energy function *U(θ)* within the round parenthesis. The *r.h.s.* term of Eq.(9) represents the power consumed by viscosity or damping. In summary, starting with the motion Eq (7) of the pendulum we have found that its potential energy *U(θ )* is a *tilted washboard potential* function of the angle $\theta$ given by:

$$U(\theta) = -\tau\,\theta - mgL\cos\theta \qquad (10)$$

plotted in Fig 4(b) for a torque *τ= 0.5* and  *mgL= 5.0*

It is now convenient, for further comprehension of the of the tilted washboard concept, to consider a simple example to be analyzed in the opposite sense: Assume now that the *tilted washboard potential energy of the physical system is known*, to then derive its motion equation. Thus, consider a particle of mass *m* that moves along a straight line with speed *v=dx/dt*, and potential energy function $U(x) = -U_0 x - U_1 \cos kx$; *k* being a proper wave number ($k = 2\pi/\lambda$, $\lambda$ the cosine period). The kinetic energy of the considered particle is $T=mv^2/2$ and therefore its Lagrangian *L= T−U* is:

$$L = \frac{1}{2}m\dot{x}^2 + U_0 x + U_1 \cos k\, x. \tag{11}$$

We may now apply the Euler-Lagrange equation, namely:

$$\frac{d}{dt}\left[\frac{\partial L}{\partial \dot{q}_i}\right] - \frac{\partial L}{\partial q_i} = 0, \tag{12}$$

where we may set $q_i = x$, to get $\frac{d}{dt}\left[\frac{\partial L}{\partial \dot{x}}\right] = m\ddot{x}$, and $\frac{\partial L}{\partial x} = -U_0 + kU_1 \sin(kx)$. Thus finally, the particle motion equation is:

$$m\ddot{x} + U_1 k \sin(kx) = U_0. \tag{13}$$

To account for a possible weak friction on the particle in its motion we might simply add a viscous-friction term $\eta\dot{x}$ and rewrite the motion equation as:

$$m\ddot{x} + \eta\dot{x} = U_0 - U_1 k \sin(kx). \tag{14}$$

Summarizing, in this section we have dealt with two mechanics examples of tilted washboard energy potentials and their corresponding dynamical equations: In the first case obtaining the tilted washboard potential function from its Newtonian motion equation Eq (7); in the second assuming the tilted washboard potential *U(x)* of a particle to be known to then derive its motion equation Eq (13).

**3 Tilted washboard potential of a superconductive Josephson junction**

In this section we shall first introduce a mathematical-physics model of a Josephson junction, and its quantum quasi-particle *phase*, to obtain its tilted potential energy function. Then we shall use this model to introduce the quantum *phase qubit* and some of its important technological applications.

**3.1 The Josephson junction washboard and its *phase quasi-particle***

The importance of teaching Josephson junctions in undergraduate physics and electronics laboratories can never be overemphasized, and that was immediately realized in the early 1970´s by research physicists and electronics engineers in industrialized countries: They began designing and constructing analogues [20, 23] of a Josephson junction using ordinary electronics components. These electronics analogues *faithfully replicate* the physics attributes and phenomenae of a real Josephson junction, with the great advantage that they *do not require cooling with liquid Helium.*

These junction analogues have been successfully are now being used by undergraduate students to perform basic and even advanced Josephson junction experiments, including for instance **(i)** the Detection of Microwaves, **(ii)** Memory cells for storing data, **(iii)** circuits to execute basic Boolean logic operations [20], and **(iv)** the generation and observation of the *Shapiro steps* [20]. The first three of these experiments are related to the application of Josephson junctions to build quantum computers as explained below (Sub-section 3.2). Moreover, Josephson junctions´ analogues are assembled with low-cost and readily available integrated circuits such as the well-known 741C and AD532. Several papers describing the construction of such junction analogues have been published *up-today* in applied physics journals [20-22] and in a physics education journal [23], so that they may be constructed by physics and electronic undergraduate students, and their instructors, across the world.

The geometrical shape of a Josephson junction (Fig. 2) clearly resembles a classical plane capacitor; two layers of metallic conductors separated by a very thin layer of an insulator (dielectric). Thus when fabricated a ***real*** junction has an intrinsic inter-electrode small *parasitic* capacitance $C$ associated to it ($C \sim 0.5$ pF). Moreover, since a real junction is never operated at exactly zero Absolute Temperature (0K), its electrodes also contain –apart from the Cooper pairs of coupled electrons of charge $2e$ – a small number of *unpaired electrons* (charge $e$) that will also cross the thin insulating layer of the junction. Thus, when at *near 0K* temperature, the junction be connected to a battery, i.e. with a voltage applied, this current of normal electrons through the junction implies that a *parasitic* parallel resistance $R$ must be also associated to the junction ($R \approx 50 \, \Omega$). Therefore, a ***real*** junction should be modelled as an ideal junction (symbol X) connected with a capacitance $C$ and with a resistance $R$ in parallel as shown in the circuit of Fig. 5; in operation this circuit is assumed to be biased with a total current $I$

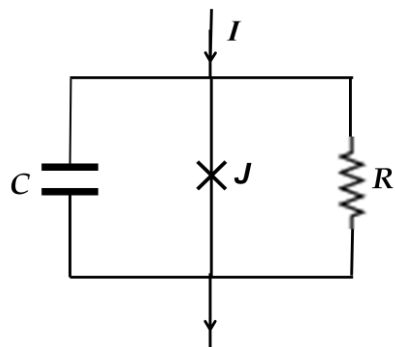

**Fig. 5** A current biased real Josephson junction $J$ (X symbol) shunted by its parasitic capacitance $C$ and its parasitic resistance $R$. $I$ is the bias current

Thus, the bias current $I$ is applied to a circuit of three branches. The current through the junction is of course the *nonlinear* Josephson super-current $I= I_c \sin \varphi$ ; nd the currents $V/R$ and $C(dV/dt)$ through the shunt resistance $R$ and shunt capacitance $C$, respectively. Kirchhoff Nodes law then allows writing a classical circuit-like equation for this real junction:

$$I = I_c \sin \varphi + C \frac{dV}{dt} + \frac{V}{R} \qquad (15)$$

This is a parametric equation of the junction, called the *McCumber-Stewart RSCJ model* (after *Resistively and Capacitatively Shunted Junction*) [24, 25] in terms of its phase $\varphi$ defined in Josephson Eq (4). This classical circuit equation we may now use to show that the junction potential energy is a tilted washboard potential function of its phase. In effect, replacing from Josephson Eq (4), the voltage $V$ into Eq (15) we may rewrite it as the following four currents equation:

$$I = I_c \sin \varphi + \frac{\hbar}{2e} C \frac{d^2\varphi}{dt^2} + \frac{\hbar}{2eR} \frac{d\varphi}{dt} \qquad (16)$$

or

$$\frac{\hbar}{2e} C \frac{d^2\varphi}{dt^2} - I + I_c \sin \varphi = -\frac{\hbar}{2eR} \frac{d\varphi}{dt} \qquad (17)$$

which is a *nonlinear* differential equation analogous to the nonlinear motion equation Eq (8) of the damped driven pendulum. By further multiplying the terms of last equation by that voltage $V$ (in Eq.(4)) we get the power consumption *(P=VI)* of the whole circuit, that is:

$$\left(\frac{\hbar}{2e}\right)^2 C \frac{d^2\varphi}{dt^2}\frac{d\varphi}{dt} - I \left(\frac{\hbar}{2e}\right)\frac{d\varphi}{dt} + I_c \left(\frac{\hbar}{2e}\right)\frac{d\varphi}{dt}\sin \varphi = -\left(\frac{\hbar}{2e}\right)^2 \frac{1}{R}\left(\frac{d\varphi}{dt}\right)^2, \qquad (18)$$

that may be suitably rearranged as

$$\frac{d}{dt}\left\{\frac{1}{2} C \left(\frac{\hbar}{2e}\right)^2 \left(\frac{d\varphi}{dt}\right)^2 + \left[\left(\frac{\hbar}{2e}\right)(-I\varphi - I_c \cos \varphi)\right]\right\} = -\left(\frac{\hbar}{2e}\right)^2 \frac{1}{R}\left(\frac{d\varphi}{dt}\right)^2 \qquad (19)$$

This non-linear differential equation may be compared with, or mapped onto, Eq (9) of the driven damped mechanical pendulum, revealing that the terms *in the curly bracket* of Eq (19) represents the sum of: *(i)* the kinetic energy of a *quasi-particle object* of coordinate $\varphi$ which "mass" $m$ would be $m = \frac{C}{2}\left(\frac{\hbar}{2e}\right)^2$, proportional to the junction capacitance $C$; plus *(ii)* a potential energy term inside the *square bracket* of the *l.h.s.* of that equation, written in terms of the bias current $I$ and the constant critical current $I_c$. This square bracket simply represents the *potential energy U($\varphi$)* of the junction quasi-particle. Note that the *r.h.s.* of Eq (19) is the analogue to the power wasted by viscous damping in the analogous mechanical pendulum of Fig (4).

Thus we may finally write the *potential energy U(φ)* of the junction quasi-particle $\varphi$ as the function:

$$U(\varphi) = \left(\frac{\hbar}{2e}\right)(-I\varphi - I_c \cos\varphi), \tag{20}$$

which is a tilted washboard potential function of *period $2\pi$*, shown in Fig. 6. This is indeed a significant result. Its *tilt is determined by the DC bias current I* value of the junction; its nonlinearity represented by the cosine term in Eq. (21), warranted by the nonlinear nature of the Josephson junction. It is the potential along which the junction quasi-particle $\varphi$ evolves along. As in the two examples in Fig. 1, the junction tilted washboard potentials, graphed in Fig. 6, have many local minimae whose depths dare given by the ratio $I/I_c$ between the bias current and the maximum super-current in the junction *RCSJ* circuit: These potential energy minimae are essential for technologically applications of the Josephson junction exploited, as shall be later explained.

Readers may have noticed that we have ended-up deriving a classical physics model (Eq (19) and Fig.6 (a)) for a quantum device, indeed a nice epistemological example. It is an example of application of the *Principle of Correspondence* of physics: In the classical limit the *phase* variable $\varphi$, of the resistively and capacitatively shunted Josephson junction, behaves as if it were a *classical* quasi-particle $\varphi$, that starting from some initial position at the top of the potential evolves nonlinearly along its tilted washboard of potential energy $U(\varphi)$ with some damping; and that may even result being trapped in one of the wells (Fig. 6(a)) to the advantage of potential important applications. For physics intuition this quasi-particle $\varphi$ reminds the motion of a mass-point moving in a gravitational field along a track having the contour of the inclined washboard [4].

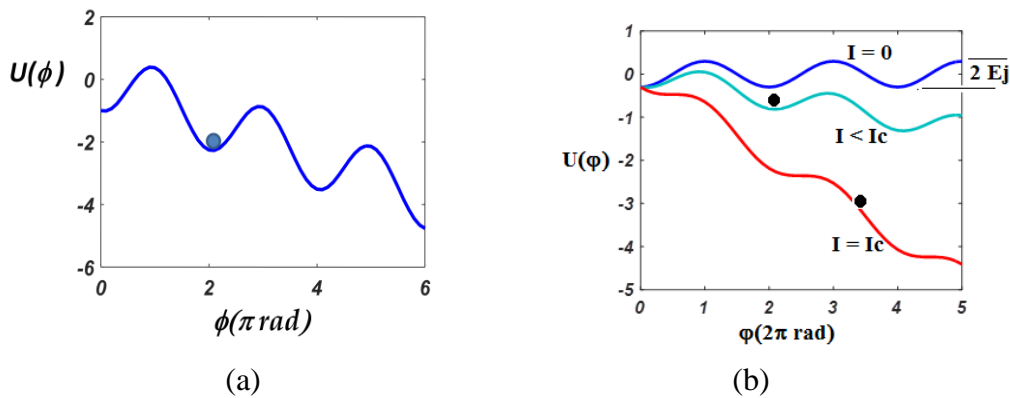

(a)                 (b)

**Fig. 6** (a) Graph of the inclined washboard potential $U(\varphi)$ of a Josephson junction, a function of period $2\pi$ of its quantum phase variable $\varphi$ with multiple minimae of equal depth; (b) Washboard potentials for three values of the bias current*: I=0; I < $I_c$* (a particle trapped: (*black dot*))*; and $I = I_c$ . $E_j$ is the Josephson energy of the junction.

The locations of the minimae of the washboard phase potential $U(\varphi)$ can be readily calculated (first derivative must be zero) to be:

$$\varphi_{min} = \sin^{-1}\left(\frac{I}{I_c}\right) + n(2\pi), \tag{21}$$

where $n \in N$ (as the washboard potential is periodic of period $2\pi$). As the phase quasi-particle of a real Josephson junction evolves down its tilted potential, with small damping, it may land at a potential well, and get confined in it, the *black dot* in Fig. 6(a), when $I<Ic$). If the bias current *I* of the junction is then increased, the washboard potential gets more inclined and the particle may eventually evolve down into a downward neighbour well. Further increasing the bias current until it equals the critical current, *i.e.* $I = I_C$, the depth of the wells will reach their infimum (as explained in the Introduction) and the phase quasi-particle will evolve down to the bottom of the potential, *i.e.* until the junction reaches a final a voltage state (recall Eq.(4)). Also note that the "*mass*" $m = \frac{C}{2}\left(\frac{\hbar}{2e}\right)^2$ of this quasi-particle remains constant since the junction shunt capacitance *C* is a constant.

It is also important to consider the energy budget of a Josephson junction when biased by a current *I*. The *constant* coefficient $E_J \equiv \frac{\hbar}{2e}I_c$ of the cosine term in Eq (19) is known as the *Josephson junction energy*: It is the *maximum e*nergy that may be stored in the junction itself. Note that it represents the amplitude of the cosine wave (see Fig. 6 (b)) of the washboard energy potential when the tilt is zero. A typical value of this Josephson energy is $\sim 10^{-18}$ J (1attoJ) for a junction current of 1mA, and it is a value fixed when the junction is fabricated. To appraise the role of a real junction in any of its application, is it is also convenient to estimate the energy confined in the capacitance *C* of the circuit *for each single electron* stored in it: It is called the *charging energy* $E_C$ of the junction and is defined as $E_C \equiv \frac{2e^2}{C}$. For a value $C \sim 1\,fF$ this charging energy is $E_c \approx$ $5 \times 10^{-24}$ *J*, giving a junction energy ratio $\frac{E_J}{E_C} \sim 10^6$.

### 3.2 Phase quasi-particle confined in a well of its tilted washboard potential

Relevant quantum phenomenae may occur when the phase quasi-particle is confined into a well of its inclined washboard potential (Fig 6 (b): There it becomes analogous to a *quantum harmonic oscillator* inside its parabolic potential $U(x)=(1/2)kx^2$ with its equally spaced eigenenergies taught in Modern Physics courses and textbooks [1-3]. But there is a major difference *w.r.t.* the quantum harmonic case: the potential wells of the tilted washboard potential of a Josephson junction are *anharmonic* (Fig. 7(a): their graphs are *not parabolic functions*, and the quasi-particle oscillations does not have *equally spaced* eigenfrequencies as a quantum harmonic quantum oscillator has. As well-known a quantum *harmonic* oscillator of frequency $v$ can only be in *proper* quantum states (called *eigenstates*) whose quantized *eigenenergy* $E_n$ values are given by the relation $E_n=(n+1/2)hv$, $n \in N$, and the energy steps between consecutive eigenenergies levels is always constant: $E_{n+1}-E_n = h\nu/2$. For the quasi-particle of the Josephson junction the *energy steps* between its consecutive eigenenergies are not equal but rather *decrease* as such energy values *increase*, (shown in Fig. 7 (a)): This is the key for the important applications of Josephson junctions. *e.g.* using it in quantum computers.

The oscillations of the phase quasi-particle, when confined in a potential well, demands that the temperature *T* of the junction has to be very low (of *mK* order), so that the thermal *Boltzmann energy* $E_B = k_B T$ (of the atoms or molecules of the junction materials) be such that $k_B T \ll h\nu_{01}$. If that thermal Boltzmann energy were greater than the oscillation energy ($h\nu_{01}$) then the oscillations of the quasi-particle may result perturbed at random, the junction not being useful at all for applications. At *T=4.2K* (liquid He) and below that Boltzmann energy is very low $E_B \sim 0.00006\ attoJ$.

### 3.3 The *phase qubit*: an important device based on the Josephson junction phase

Conventional computers rely on electronic circuits of semi-conductor components designed to perform arithmetic and logic operations analogically. These circuits operate with data represented in a discrete basis of numbers, *e.g.* the *integer* binary base {0,1} corresponding to two discrete voltage values of an electronic component in the motherboard circuit, that is called a *bit*. As well known our conventional computers require truly large numbers of bits, to execute our computing instructions.

By the 1980´s it became clear that such conventional computers cannot solve rather complex problems of the quantum world, such as simulating a chemical reaction between two organic molecules, or even simpler to simulate *the motion of a gas of neutral atoms* at temperatures close to 0K. That realization finally led to the conceptual ideas [26] of a computer based on *quantum objects* capable of providing large number of quantum states to do computations, plus large data storage, *e.g.* the phase quasi-particles of Josephson junctions, photons in a laser beam, ions trapped in crystals. One each of such quantum objects was named *quantum bit*, a term abbreviated as *qubit.*

A *qubit* is defined as a *quantum object with* a well defined *basis of two distinguishable quantum eigenstates*; a basis that can be used to generate a large number of additional combined quantum states (by a simple linear combination of the basis eigenstates). Because of its simplicity of operation, the first type of qubit studied by 1984 was based on a Josephson junction: the two required well-defined states being two quantum states of energies, say *E₁* and *E₂*, of its quasi-particle phase *φ* when confined in an anharmonic well, Fig. 7(a).

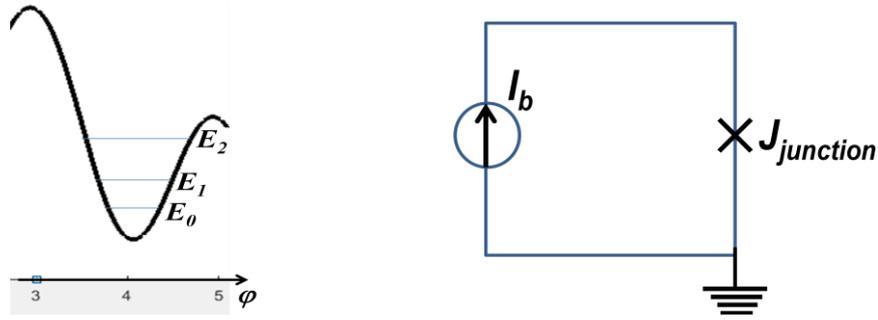

**Fig. 7** (a) Quantized energy levels $E_1$, $E_2$, $E_3$ of the phase quasi-particle, when confined in an *anharmonic* well of its tilted washboard potential; (b) Phase qubit circuit and its bias current.

Therefore they are called *phase qubits*. Apart from the phase qubit, many other circuit configurations for qubits are being used today to build quantum computers, the simplest of them being based on circuits where the key component is a Josephson junction. The three simplest qubits based on a single Josephson junction are: the *Phase qubits*, the *Charge qubits* and the *Flux qubits*, that correspond to three different circuit configurations. In what follows we shall only consider the *phase qubits* since this is the only kind of qubit with a potential energy function that plots as a tilted washboard potential.

Why a *phase qubit*? A phase qubit is technically implemented with a Josephson junction of large Josephson energy $E_C$, so that $E_J \gg E_C$, cooled down to millikelvin temperatures. A large $Ej$ value means that its phase $\varphi$ is well-defined and may then be reliably used as the variable to represents its quantum state. As shown in Fig. 7(b) the circuit of a phase qubit is very simple: a single external current source is required, to provide the *bias current* $I_b$ to the junction. It can be operated with bias currents of mA order that may be delivered by ordinary lab current generators, which is technically advantageous. Other type of qubits may demand much smaller bias currents, requiring therefore specially designed current generators.

In the symbolic vector notation invented by P. M. Dirac [1, 2] for quantum states, each *quantum state* of the quasi-particle $\varphi$ is denoted as $|\varphi\rangle$: "the phase symbol $\varphi$ enclosed inside a *vector symbol* $|\ \rangle$" which is called a *ket*. It is only a convenient short-hand notation used in quantum mechanics. Any vector *ket* (quantum state) of a quantum object belongs to a vectorial space of quantum states.

Consider the phase quasi-particle to be initially in state $|\varphi_1\rangle$ of energy $E_1$, when confined in an anharmonic potential well (Fig. 7 (a)). By irradiating the junction with a *microwave radiation pulse* of the required and accurate frequency $\nu_{21} = (E_2 - E_1)/h$, the phase particle will be exactly excited from state $|\varphi_1\rangle$ to its next higher state $|\varphi_2\rangle$ of energy $E_2$: these two *well-defined quantum* states then satisfy the required condition for the junction to be used as a computer *qubit*. Had the quasi-particle $\varphi$ been in the initial quantum state, of the $|\varphi_0\rangle$, of energy $E_0$ (Fig. 6), that accurate microwave pulse could not had produced any quantum transition simply because $\nu_{21} \neq \nu_{10}$ (recall that *consecutive* quantized *energy levels in any anharmonic potential are not equally spaced*). In passing, technologies to generate microwaves pulses of extremely accurate frequency values, is fortunately available since decades ago.

In the case of the *phase qubit* the quantum energy state $|\varphi_1\rangle$ of energy $E_1$ may be assigned to represent the integer value 0, while the state $|\varphi_2\rangle$ of energy $E_2$ may represent the value 1. A vector space of basis $\{|\varphi_1\rangle, |\varphi_2\rangle\}$ is thus obtained for computing operation. Now, if we fabricate a quantum computer with *4 phase qubits*, the corresponding *vector space* has the following *basis* of $2^4=16$ *vectors*, each of them being the product of four kets:

$$\{ |0,0,0,0\rangle, \ |1,0,0,0\rangle, \ |0,1,0,0\rangle ... |1,1,1,0\rangle, \ |1,1,1,1\rangle\}, \qquad (22)$$

each of these obtained by *orderly writing* all possible products of the four qubits states, each one either in state $|0\rangle$ or in state $|1\rangle$. For comparison, the equivalent conventional electronic computer would require to be assembled with at least 16 electronic *bits*, not just four. Therefore a quantum computer with only *15 qubits* corresponds to a conventional computer with at least $2^{15} = 32768$ *electronic bits*!

But wait, using a quantum computer it is even better than that! Because according to the *Principle of Superposition* of quantum mechanics [1-3] a single qubit, whose basis is $\{|\varphi_1\rangle, |\varphi_2\rangle\}$, can also be *in infinite states* that are linear combinations of the form: $|\varphi\rangle = \alpha |\varphi_1\rangle + \beta |\varphi_2\rangle$, where $\alpha$ and $\beta$ are a pair of *arbitrary complex numbers* that should only satisfy the condition $|\alpha|^2 + |\beta|^2 = 1$. For instance, choosing the complex numbers $\alpha = 2$, and $\beta = \sqrt{3}\,i$ gives a new qubit state:

$$|\varphi\rangle = 2|\varphi_1\rangle + \sqrt{3}\,i\,|\varphi_2\rangle, \text{ since } 2^2 + (\sqrt{3}\,i)^2 = 1 \qquad (23)$$

So, the *16 quantum states* of the vector basis of four qubits, in the set (22) written above, can be linearly combined to generate a huge large number of quantum states, all available for doing complex computations, such as simulating a complex phenomenon e.g. controlling by simulation the action of an organic enzyme in a given human biological process, or the emblematic mathematical *travelling salesman* problem.

Apart from being capable of storing data as memory qubit states, a quantum computer should also provide means to implement the basic logic gates (of Boolean algebra: OR, AND, NOR and the like), plus implementing sophisticated computing algorithms, *e.g.* an algorithm to obtain the Fourier transform of a musical sound [29]. Yet, this work, devoted to show the importance of the tilted washboard potential energy of Josephson junctions and electro-optical systems, we cannot go any further with the quantum computation subject. Reports and books [29-31] on quantum computation have been published since 2000; that show the important role of Josephson junctions for fabricating qubits for the circuits of quantum computers, and explaining all the implicit technicalities, *e.g.* how to protect the qubits from external electro-magnetic perturbations. Today (2023) quantum computers with several tens of qubits are freely available for implementing very useful algorithms *e.g.* the quantum cryptography ones.

## 4. Electro-Optical potentials: Bessel beams potentials and Standing wave potentials

This is an introduction to some physics methods used for the micro-manipulation, diffusion transport, and trapping of microscopic and sub-microscopic dielectric particles, say *biological cells* and *neutral atoms*; methods that are based on the interaction of such dielectric particles with washboard-like *electro-optical* potentials generated with laser beams. Fortunately, the generation of such energy potentials happens to be within the reach of undergraduate physics and engineering students.

As mentioned in the Introduction when a tiny *dielectric* particle is placed in a focused region of a laser beam of electric field $E(r, \lambda)$, wavelength $\lambda$, angular frequency $\omega = c/(2\pi\lambda)$, it does acquire an induced electric dipole moment $d$ proportional to its particle polarizabilty $\alpha(\lambda)$. Such electric dipole then interacts with the electric field to acquire a quasi-periodic *electro-optical potential* $U(r) \propto |E(r)|^2$ (Eq (6)). What is most important is that the dielectric particle may result confined, or forced to evolve along this potential of energy, by the same electric field that created it. Analogously to the trapping of the Josephson junction phase quasi-particle $\varphi$ inside a well of its washboard potential $U(\varphi)$, the dielectric particle may also end up being confined in one of the wells of its electro-optical potential $U(r)$. Once trapped in a well, if it is a neutral atom, it is possible to excite it to make a quantum transitions between its quantized Bohr atomic states. In this section we shall refer to two kinds of periodic electro-optical potentials generated with lasers, namely:

**(i)** The washboard-like periodic potentials generated with the so-called laser *Bessel beams*,
**(ii)** The Standing Wave potentials generated by two counter propagating laser beams.

Since an electro-optical potential is proportional to the intensity of the field that generates it, it is common to refer to the potential in terms of the field; when necessary one must make the distinction between the physical magnitudes.

### 4.1 Electro-optical washboards potentials generated with Bessel beams

As mentioned in the Introduction most laser beams used in undergraduate optics experiments and lecture demonstrations have transverse irradiance profiles given by a *Gaussian*: $I(r) = I_0 \exp[-2r^2/w^2]$ function, $r$ being the transverse radial coordinate, $w$ the beam waist (Fig 3(a)). Interestingly, starting with a well-collimated *Gaussian* laser beam it is possible to generate a new kind of laser beam with a different transverse intensity distribution. Thus, in most of the applications presented below, the laser beam used for trapping particles, has a transverse *quasi-periodic* amplitude profile represented by a *Bessel function*. A *zeroth order Bessel beam* has a transverse amplitude given by the $J_0(kr)$ Bessel's function (Fig. 8(a)), whose transverse intensity, observable on a screen, is a pattern of concentric rings of decreasing light intensity (Fig. 8(b) represented by the function $J_0^2(kr)$, $k=2\pi/\lambda$ being the wave number, $\lambda$ its wavelength, usually in the *I-R* region of the Electro-Magnetic spectrum.

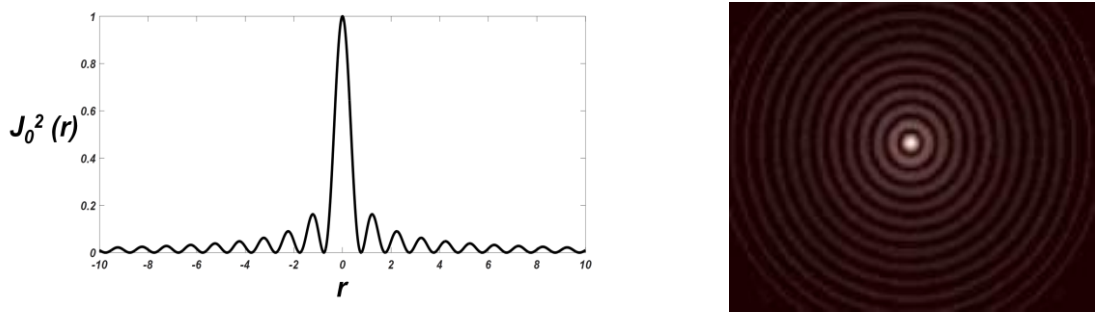

**Fig. 8** (a) Mathematical plot of the transverse radial amplitude of an *ideal* 0$^{th}$ order Bessel beam; (b) its transverse intensity pattern of concentric rings, when observed on a screen.

In 1954 E. McLeod reported [32] that: "a *plano-conical* lens − baptized *Axicon* by him− may be used to generate a *Bessel* $J_0$ beam by irradiating it with a *Gaussian* beam". As shown in Fig 9 (a), an axicon is an optical element with a flat *left* face, its *right face* being conically shaped surface, of rather large vertex angle, say ~170 °. Axicons can be made of optical glass (say *BK7*), Plexiglas, fused silica, or any other optical material. Figure 8(b) illustrates the typical and simple McLeod experimental set-up to generate a Bessel beam using an axicon along its optical axis: A well-collimated Gaussian laser beam of radius *R* is incident from the left onto the axicon along its optical axis. To generate *a* strong enough *dipole electro-optical potentials*, this beam must be of sufficient high intensity (say ~1 W).

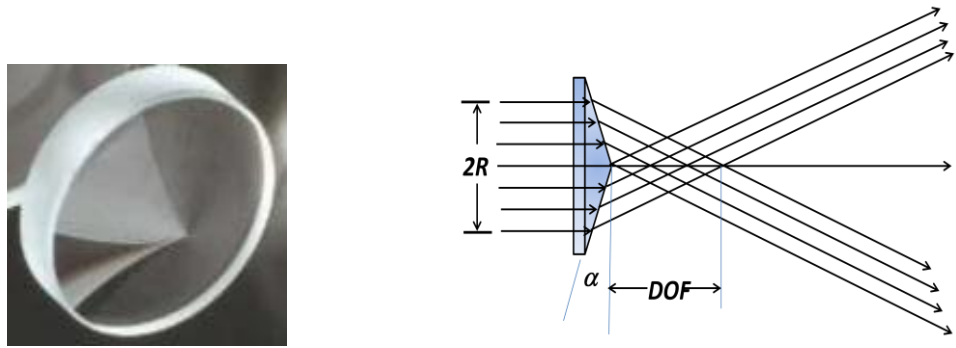

**Fig. 9 (a)** An axicon; **(b)** Optical set-up to generate a zeroth order Bessel beam from a well collimated Gaussian laser beam (radius *R*) incident onto an axicon of large vertex angle $\beta = (180° -2\alpha )$, $\alpha$~ 5°. A Bessel beam is generated in the DOF space region were the two beams superpose.

Due to its conical face the axicon refracts the incident laser beam *plane wavefronts* as coherent *conical* wavefronts that converge towards the optics axis of the axicon (Fig 8(b)). These conical wavefronts superpose, and *interfere*, in the space region to the right of the axicon - about its optical axis- whose length is the so-called *depth of field (DOF)* of the axicon [32, 33]. As predicted by McLeod, the conical waves from the axicon interfere to generate a light beam propagating along the *DOF* region and whose *transverse* radial intensity distribution approximates a zeroth-order *Bessel laser beam* (Fig. 8 (a))

That *DOF* region is precisely the interaction region where a transverse Bessel washboard-like electro-optical potential *U(r)* is generated – and has been exploited in multiple applications developed since about 40 years ago–when tiny dielectric particles are deposited there, and the laser Bessel beam induces in them electrical dipoles. It is rather important to be aware that the intensity along a *longitudinal* cross section of a Bessel beam has a *fringe-like pattern* profile (shown in Fig 10), totally different to its transverse radial structure (Fig. 7 (b))

Provided an axicon be available the simple optical set-up to generate a Bessel laser beam (Fig 9(b)), may be assembled by undergraduate students. An axicon offers to us an additional outstanding diffraction phenomenon. In effect, if the generated Bessel beam is obstructed placing a very small opaque obstacle across it -on the axicon axis- then the Bessel´s beam diffracts about the obstacle and reconstructs itself to the right of the obstacle along the axicon axis, *as if no obstacle was inserted*. This unexpected optics phenomenon has been baptized *immunity to diffraction* [10, 17], and can be exploited in applications. In passing, it is interesting to mention that axicons –apart from the generation of electro-optical potentials –have a number of other applications in physics, medicine and engineering namely: laser corneal surgery, Calcium imaging of neural cells, laser drilling, and the building of solar concentrators.

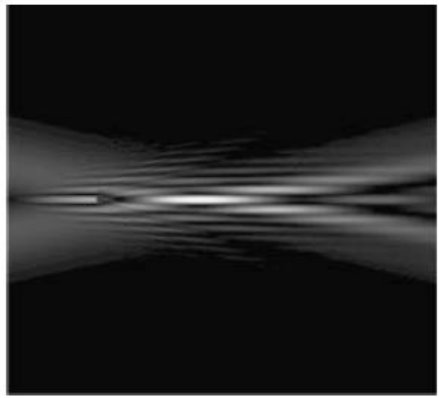

**Fig. 10** Longitudinal cross-section of a Bessel beam along the DOF region of an axicon, showing its fringe-like pattern; also showing the beam self-reconstruction after the small opaque object placed on the axicon axis (after Mc Gloine *et al* [10])

The transverse *radial* electric field electro-optics potential $U(r) \propto |\mathbf{E}(r)|^2$ of a *zeroth-order Bessel´s beam* obtained with an axicon, is quasi-periodic and rather similar to a symmetric tilted washboard potential (Fig. 8(a)): the high peak of potential energy at the centre of the intensity pattern, and *wings* with small tilts, at the sides. As shown in Fig 9(b) a laser beam, whose transverse intensity profile is Gaussian (Fig. 3(a)) is used to generate the Bessel beam. Unavoidably, the Gaussian light intensity distribution of the plane wavefronts –that are refracted by the axicon– introduces a *residual* parabolic-like potential that superimpose on the electro-optics potential created by the Bessel beam $J_o^2$ intensity [10]. Thus clearly observable tilts of the electro-optical potential are naturally generated in the research experiments with such Bessel´s beams (Fig. 11), and are exploited in research and applications [10].

The work of Tatarkova *et al* [6] is a good example of the application of a Bessel beam electro-optical potential as a *particle motor*. The Bessel laser beam has a wavelength of ~1μm and the wells of the washboard electro-optical potential generated were ~5μm wide. The aim was to study the Brownian motion of microscopic dielectric silica microspheres, of diameter ~1μm. The particles were dispersed in water contained in a small cell, at very low concentration (~0.2 %) so that interactions between the particles could be neglected. The trapping and motion of the particles in the tilted potential were then observed (illustrated in Fig 11). A typical low-cost microscope objective (say 20X, NA=0.4) was used to observe the particles moving in the tilted washboard potential. In their work the researchers used a slightly converging irradiating Gaussian beam to generate a pronounced tilt (Fig. 11) in the washboard electro-optics potential given by an axicon and collect the particles at the central well of the potential.

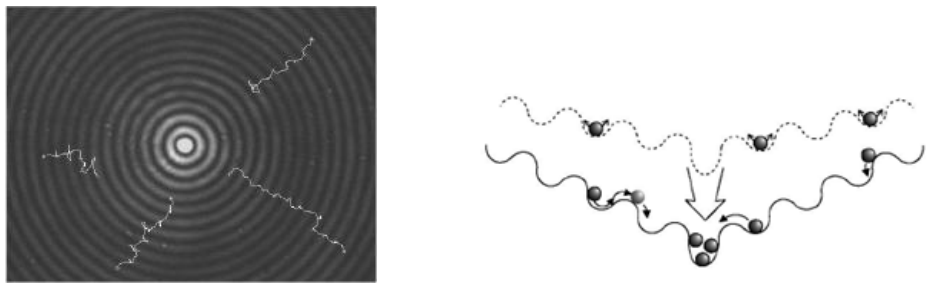

**Fig. 11** Evolution of micrometre size particles in the tilted transverse cross-section of an electro-optics laser Bessel electro-optical potential: (a) particles trapped across wells of the potential; (b) due to the tilted symmetry the tiny particles may run down to the central well (from Tatarkova *et al* [6]) to be channelled out by the higher radiation pressure of the electric field there.

Light interference and light diffraction are subjects of undergraduate physics courses. For instance, the well-known Fraunhofer diffraction pattern of a thin slit represented by a *sinc² function* [34], and notably the appealing ringed diffraction pattern of a circular aperture [13, 16], whose amplitude is given by the function $J_1(ar)/a$ [16, 34]. However, it must be accepted, that the mathematical-physics models [10, 32, 38] that explain the generation of a Bessel´s beam, with an axicon, are beyond the scope of most undergraduate optics courses. Yet, undergraduate students may easily assemble the experimental set-up required to generate those beams, thus beginning to understand this advanced cases of diffraction and their applications.

Fortunately, there is a simpler and low-cost alternative –discovered by J. Durnin in 1987 [36] – for generating a $J_0$ Bessel beam without an axicon. This alternative optical set-up is within easy reach of undergraduate physics and engineering students. In effect, as explained in [17, 36] a simpler way to generate a $J_0$ Bessel beam is to diffract a Gaussian laser beam with a *ring-slit* aperture (*i.e.* a *circular-slit* aperture), in a simple low-cost experimental set-up that can be readily implemented by undergraduate students (Fig. 12). It consists of a well collimated Gaussian laser beam incident on a *ring-slit aperture* placed at the back focal plane of a positive lens of focal length $f$, that then refracts the light beams diffracted by the ring-slit aperture.

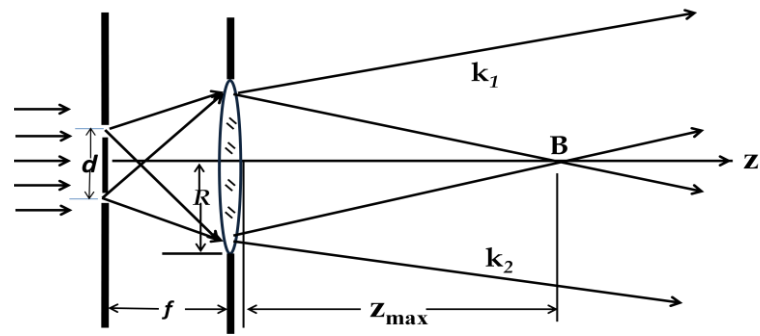

**Fig. 12** A well collimated *Gaussian* laser beam irradiates a narrow *ring-slit* aperture of diameter $d$ and slit width $\sim d/10$, placed at the back focal plane of a positive lens (focal length $f$): A laser Bessel beam is then generated in the near diffraction region of length $z_{max}$ were the two refracted beams, of vector direction $\mathbf{k_1}$ and $\mathbf{k_2}$ superpose.

Students can manufacture a ring-slit aperture starting with a thin black ring, plotted it with a computer program or simply photographed. This ring may be frame-grabbed with the high-resolution camera of a present mobile phone, and the picture processed (*e.g.* size reduced, contrast reversed, pixels resolution augmented) using a *free-download* image processor software such as the excellent *Image J* [37]. Finally, the ring-slit aperture may be printed in a slide, as a high clear annulus of few millimeters radius, and narrow with, on an opaque background.

By placing this ring-slit aperture at the back focal plane of the positive lens (Fig 12), and irradiating it with a collimated Gaussian laser beam from, say a 6mW diode laser (wavelength $\lambda= 670$nm) the superposition of the diffracted waves from all points of the circular slit renders beams of conical wavefronts that creates a Bessel beam at the right of the lens; in the near spatial region of length $z_{max}$ about the optical axis (Fig. 12) between the point B and the lens (see [17, 38] for more details)., As explained in the work of Grimm et al [37, 38], this Bessel beam can be derived as the Fourier transform of the irradiated annular slit. In Fig (12) the inclined light beams along direction $k_1$ and $k_2$ represent the circular-slit diffracted waves just on the plane of that figure.

It may be shown that the superposition, of the beams diffracted by the annulus aperture (Fig. 12), give the expected Bessel beam propagating about the lens axis, and that the transverse electric field intensity in the focused region of the lens of length $z_{max=}=2Rf/d$, along the aperture $z$-$axis$, is that of a Bessel $J_0$ beam and given by [38]

$$I(r,z) = I_0 \left(\frac{z}{z_{\max}}\right) exp\left[-\frac{2z^2}{z_{max}^2}\right] J_0^2(k_r\, r). \tag{24}$$

$k_r = (2\pi/\lambda)\, \sin\, \theta$, and $\theta$ any azimuthal angle (in passing, this beam irradiance is incorrectly written in some published papers). Again, as in the case of the axicon, the *transverse radial irradiance* of the Bessel laser beam is proportional to $J_0^2(kr)$, and appears graphed in Fig 13, with an increased tilt added by the incident Gaussian beam intensity profile: (a) at 100 mm from the lens, (b) at 400 mm from the lens.

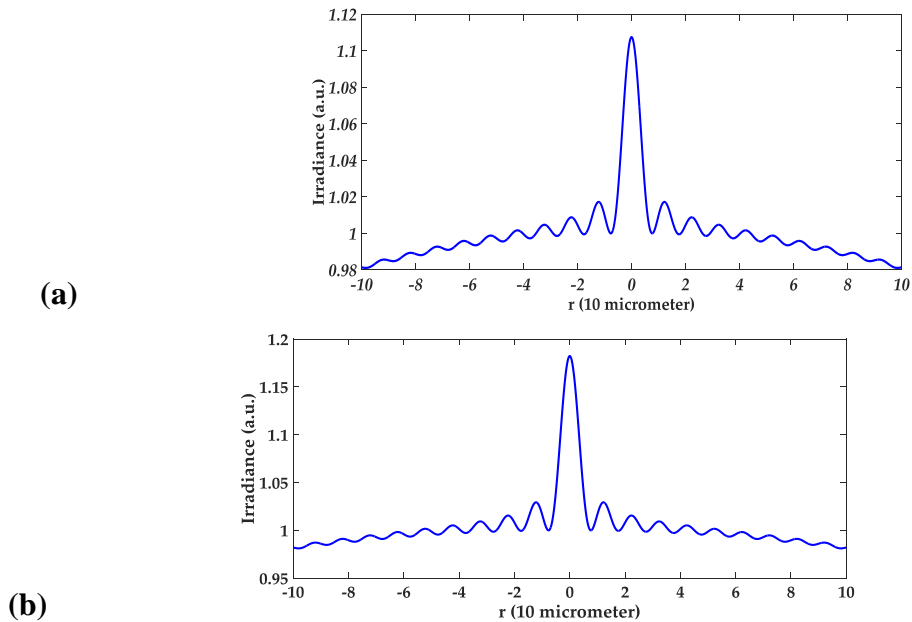

(a)

(b)

**Fig. 13** Graphs of the transverse radial irradiance of a Bessel beam when generated by a a circular-slit aperture (of $z_{max}=700\ mm$) at: (a) $z=100\ mm$, (b) *400 mm*. Graphs obtained adding a Gaussian intensity bias to the *Bessel* intensity relation given in Eq. (25).

It is important to emphasise that the *transverse irradiance distribution* of a Bessel beam generated in the laboratory with either with a circular-slit aperture (or with an axicon) is necessarily tilted [17, Fig.3]. The tilting is caused by the Gaussian intensity profile of the light wavefronts used to irradiate the circular-slit aperture that produces an *indelible residual* intensity that tilts the laser Bessel profile irradiance as shown in Fig. 13 above (*c.f.* actual experimental Figs. 3 in [17]). One must realize that when Eq. (25) is used to graph the transverse profile of a generated Bessel beam one is assuming that the collimated irradiating beam (of Fig. 12) has *ideal* wavefronts of *constant transverse irradiance*, which is *not true for an illuminating Gaussian beam*. The Gaussian irradiance profile leads to a weak symmetrical potential superimposed onto the ideal Bessel profile and tilt it. Thus the graphs in Fig.13 incorporate this residual parabolic-like potential due to the Gaussian beam and resembles actual experimental plots obtained in experiments [17]. Again, that tilting may be even increased using *a slightly converging* Gaussian beam, instead of a *well collimated* beam (as done by Tatarkova *et al* [6] in their work); thus producing a tilted transverse intensity distribution beam, that better resemble the profile of a tilted washboard electro-optical potential function (Fig. 11).

In the last two decades a significant number of research work and applications that exploit Bessel beams have been developed, including among others: the interaction of these beams with dielectric particles and their optical manipulation [10, 39], *In Vivo* imaging of neural cells with alkaline-earth ions [40], trapping and motion of of long and thin bacteria such as *Escherichia coli* [6], molecular motors [41], transfer of orbital angular momentum to low-index particles [42], this last reference in a physics education conference proceedings. The great majority of these applications and research experiments require additional setups illumination, and for proper observation (e.g. microscope objectives, telescope optics) and for accurately controlling the particles motion along the tilted washboard potentials. Necessarily, these additional set-ups lie beyond the scope of this work, and have to be studied in the respective referenced papers.

It must be mentioned that apart from the trapping of particles in Bessel beams generated potentials, and standing wave potentials, other interesting energy potentials can be generated, notably the *ratchet and pawl potential* conceived by R. P. Feynman back in 1963 [9, Vol 1, p46] to use it as a Brownian particles motor, experimentally implemented and reported in 2010, by two research groups. Also, apart from Bessel beams, other laser beams of different transverse irradiance have been used to motorize cold gases of atoms, such is the case of the Bessel-Laguerre beams mentioned in [10].

### 4.2 Standing wave electro-optical potentials: *Dipole Optical Lattices*

In the introduction we mentioned standing waves of sound. Such kind of waves, particularly those in strings instruments, is a subject treated in *Waves* textbooks [43, 44]. Below we consider analogous light standing waves of electro-optical potentials, generated with lasers, and introduce some of their outstanding applications when trapping sub-microscopic particles, even neutral atoms! Let us recall the simple set-up of the Wiener standing wave [34]: in which a well collimated coherent light beam of wavelength $\lambda$ is incident onto a flat mirror along its normal (Fig.14). The incident beam and the *counter-propagating* reflected beam generate a light standing wave. It is a spatially periodic interference pattern, with *antinode* planes of maximum intensity of period $\lambda/2$. Using a He-Ne collimated laser beam that period is rather small ~0.32µm. (much less than the diameter of a red cell) Good examples, of counter-propagating light beams creating standing waves, occur in both arms of a Michelson interferometer [13, 34, 44] a useful interferometer studied in undergraduate physics laboratories across the world.

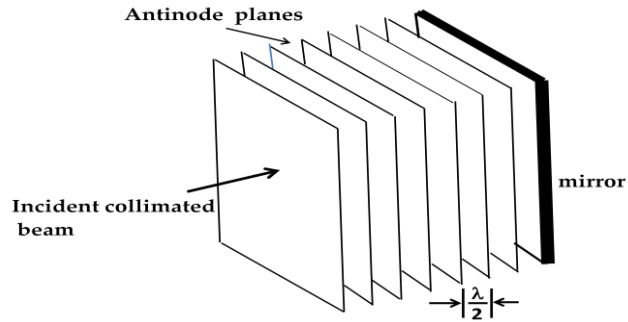

**Fig. 14** A collimated coherent beam incident along the normal of a plane mirror generates with the counter-propagated reflected beam a standing wave: the highest intensity and periodic *antinode* wavefront planes are separated a distance $\lambda/2$.

Clearly a laser standing wave is indeed a far easier way for creating a periodic electro-optical potential [17] than using a Bessel beam; they are also normally generated with Gaussian laser beams. Moreover, in a laser standing wave (Fig. 14 (a)) the depth of the potential wells can be easily decreased or increased by the experimentalists by varying the laser beam intensity (*e.g.* with two crossed polarizers). It is well-known [13,34,43,44] that a 1-dim standing wave has an amplitude proportional to the function $sin(kx)$, where $k = 2\pi/\lambda$ is the wave-number of the wave, $\lambda$ the wavelength. In the case of a laser light standing wave interacting with a particle electrical dipole, the 1-dim electro-optics potential energy $U(x)$ acquired by the particle may be written as:

$$U(x) = U_0 \sin^2(kx), \qquad (26)$$

where $U_0$ is proportional to the maximum intensity $I_l$ of the laser and to the induced particle polarizability $\alpha(\lambda)$ for the laser wavelength [46]:

$$U_0 = -\frac{Re(\alpha(\lambda))}{2c\,\varepsilon_0} I_l \qquad (27)$$

A standing wave electro-optical potential is plotted in Fig. 14 (a), also showing a few particles, say *neutral* atoms, confined in the lattice potential wells. The whole geometry of these electro-optical potentials can be varied: for instance the period of the standing wave may be varied by varying the angle between the interfering laser beams that creates it. In practice the profiles of laser generated standing waves are always symmetrically tilted. Since the usual laser beams that create these standing wave electro-optical potential have Gaussian intensity profiles, an indelible weak harmonic confinement is superimposed on the *sin²* periodical lattice [18], as was explained in Sub-section 4.1.Thus the actual particle confinement in experiments is always inhomogeneous as shown in Fig.15 (b); the profile of these standing wave potentials have *indelible tilts*: they are *symmetrical* tilted washboard potentials, a feature that is favourably exploited in applications.

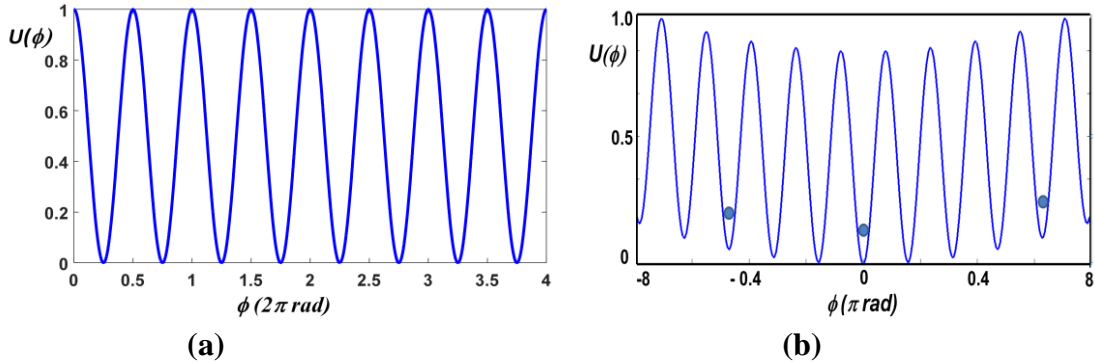

**(a)** **(b)**

**Fig. 15** Plot of an ideal laser standing wave periodic electro-optical potential illustrating tiny dielectric particles in the potential wells (b) Actual tilted standing wave pattern generated by two counter propagating Gaussian laser beams: indelible tilts are imposed by the Gaussian intensity profile of the beam that creates the standing wave..

The laser beams used in these standing-wave electro-optical potentials have powers of order ~ 0.5 W, or higher, in order to warrant induced electric dipole moments strong enough for strong electro-optical forces acting on the particles. The typical laser used emits visible or near-IR light, say λ= 1.064 μm, therefore the width of the wells is of order $\lambda/2 \sim 400$ to $800 nm$. A typical application of these potentials, in the last two decades and the present decade, is the trapping of neutral atoms, say K or Sr atoms, in the standing wave wells to create advanced technological devices, as explained below.

The periodic wells of a 1-dim laser standing wave trap dielectric particle dipoles in a 1-dim periodical array called *dipole optical lattice* (Fig 15), or simply *optical lattice*; names arising from their similitude with the case of atoms trapped in a crystal (*e.g.* in the cubic unit crystal cell of *NaCl*, the $Na^+$ and $Cl^-$ ions are stacked at fixed positions in the cell [13]). Thus, these laser light optical lattices can be seen as if they were *artificial light crystals.* Moreover, it is not difficult to produce 2-dim, and 3-dim orthogonal laser standing waves, superimposing two or three pairs of counter-propagating lasers beams, respectively, in orthogonal directions. These higher dimension optical lattices then generated is a 2-dim, or a 3-dim periodic electro-optic potential, that can be used to trap atoms on fixed positions along a plane, or in 3-dim space, respectively [18].

An important application of standing wave electro-optical potentials is the investigation of *very low density* ensembles of neutral atoms held at low temperatures of 1mK order, called *ultra-cold gases of atoms*. For instance, a gas of K atoms at a density equivalent to $1/10^6$ of the ambient air density (*i.e.* only ~$3\times10^{14}$ K atoms/cm$^3$) enclosed in a cryogenic vessel at millikelvin temperatures. Such diluted ensemble of particles is like a tenuous cloud of atoms, so widely separated that need to be modelled as quantum objects, and represented as *quantum wave packets* of quantum mechanics [1,2], not as tiny isolated classical mechanics objects. When an ultra-cold gas of atoms interacts with a laser optical lattice, *thousands of atoms* may become orderly trapped in the lattice (in some cases at the high intensity antinodes because the confining electrical force on electrical dipoles is proportional to field intensity). The resulting array of atoms is a robust and *totally controllable* quantum system for the experimentalists to do research *e.g.* quantum simulations of complex systems. As already mentioned above, the depth and even the geometry of the optical lattice are under their complete control. An obliged expectative of these optical manipulations of atoms and molecules, is that they should be as easy for the experimentalists to manipulate, as when they manipulate light beams with conventional optics (lenses, polarizers, prisms, etc). The entire geometry of these trapping potentials is at the hand of the experimentalists, *e.g.* by interfering lasers of different transverse intensities such as Bessel-Laguerre laser beams. If the angle of superposition of the counter-propagating laser beams is continuously variable an *accordion optical lattice* is obtained [47] offering additional possibilities for research and applications (*e.g.* in quantum computing with neutral atoms)

Optical lattices have been used since about year 2000 to investigate ultra-cold quantum gases of either *bosons or fermions* [10]. These quantum particles trapped in optical lattices can be considered as simulators of the quantum world, as R. P. Feynman originally conceived for a quantum computer [26, 29]: "a powerful simulator in which a highly controllable quantum system can be used to simulate the dynamical behaviour of another complex quantum system". Optical lattices are also finding applications in quantum optics, quantum information processing, and for testing condensed matter theories. A good account of these applications may be found in Bloch´s paper [18] and references there in.

Dipole optical lattices are the key element of the most accurate atomic clocks implemented up-today. In this key application neutral atoms (say of isotope $^{87}$Sr) are trapped in an optical lattice created with a laser of some wavelength $\lambda$. By irradiating the confined atoms with microwave pulses of the proper frequency $v_{12}$ a quantum transition, from atomic state |$S_1$> to a higher energy state |$S_2$> should likely occurs. Such microwave pulses of the required frequency $v_{12}$ can be accurately generated with present microwave technology. Yet, usually there arises a serious experimental perturbation. As mentioned in the Introduction the induced electrical *polarizabilities* $\alpha_1$ and $\alpha_2$ – of two atomic quantum states |$S_1$> and |$S_2$> – are two *different functions of the laser wavelength λ used to create the trapping optical lattice*: the induced electric dipoles of these two states are different. This difference is a significant perturbation for the expected atomic transition to occur, and it does not occur at the expected frequency $v_{12}$, *i.e.* the frequency of the microwave pulse has to be shifted, by the experimenters, to a different value for the transition to occur.

In 2003 H. Katori made a significant discovery: While investigating whether it was possible to create an optical lattice with a laser of such a wavelength that the two polarizabilities $\alpha_1(\lambda)$ and $\alpha_2(\lambda)$ were equal, thus avoiding the perturbation, and therefore *no shift* in the required transition frequency; using the mathematical-physics model of atoms polarizability he was able to show [48] that such wavelength does exist (see Fig, 16) at least for some atoms, *e.g.* for some isotopes of Sr and K. It is called the *magic wavelength*. After three years of experimental research it was finally possible for . Katori *et al*, to create the proper optical lattice using a laser of a wavelength $\lambda_{Magic}$, called the *magic* wavelength, and to observe in the laboratory the transition between two quantum atomic states $|S_1\rangle$ and $|S_2\rangle$ occurring at the expected frequency $\nu_{12}$, with no shift at all.

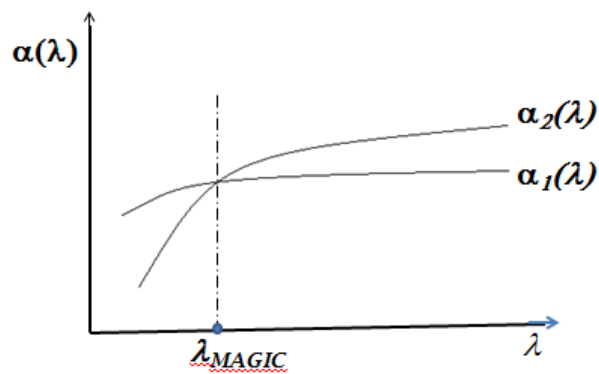

**Fig. 16** Dipole polarizabilities $\alpha_1(\lambda)$, $\alpha_2(\lambda)$ of two atom quantum states, $|S_1\rangle$ and $|S_2\rangle$, plotted *vs.* wavelength ($\lambda$) of the laser that creates an optical lattice to trap the atom: the two polarizabilities are equal at the magic wavelength $\lambda_{Magic}$.

Today we know that for atoms of the isotope $^{88}$Sr the *magic wavelength* for its two atomic states $5S_0$ and $5P_1$ is 473.371(6) nm. In the case of Yterbium atoms (Y), the magic wavelength for the transition between its states $6S_0$ and $6P_1$ has been measured to be 1035.68(4) nm. In passing, the mathematical-physics models of *electric dipole polarizability* of atoms and molecules are presented in quantum mechanics textbooks for physics majors, *e.g.* [3, Vol. I].

Without perturbation the required microwave absorption frequency between two quantum states of the atoms trapped in an optical lattice created usin the *magic* wavelength , can then be determined and generated extremely accurately. For instance, today such frequency may be measured with accuracies up to 18th decimal places, that is one second error in 3000 million years!, providing the possibility of fabricating the most accurate *optical atomic clocks* thus far conceived. Portable optical atomic clocks, based on optical lattices, are now in the market (2023) with an accuracy of 1 second lost or gained in 50000 million years!

Atomic optical lattice setups are so reliable and accurate that they are being used to probe small differences of Earth gravitational acceleration with great accuracy. Such optical atomic clocks are even being considered for gravitational waves detection [49], thus far only detected with large interferometers, *e.g.* LIGO at Livingstone USA (a Michelson type interferometer of 4.8 km long arms). By 2013, two optical lattice clocks containing about 10000 atoms of isotope $^{87}$Sr already stayed in synchrony with each other with a precision of at least $1.5 \times 10^{-16}$.

## 5. Discussion and Conclusions

When teaching natural sciences, even a formal science, it is valuable to count with resources that can provide unity of conceptualization and enhanced comprehension, for students to learn faster, their cognition lasting more. An example is to apply Dimensional Analysis to physics models and results, *e.g.* as done with our Eq (19) when a "mass" was assigned to the coefficient of the first term of a mathematical relation derived in the Josephson junction model. In this work we have presented another teaching resource, certainly a useful one: energy potentials whose graphs are like tilted washboard profiles. As shown, these washboard potentials are recurrent in the analytical models of important classical and quantum physics systems, notably in the model of a real Josephson junction; and later when we presented laser beams of *Besselian* intensity profile, whose interacting with dielectric particles generate electro-optic energy potentials with washboard profiles that may be easily tilted by the experimentalist.

As quoted in the Introduction, washboard potentials are "ubiquitous seminal landscapes across diverse physics researches and technologies, at present receiving significant attention" [8]. Notwithstanding, washboard energy potentials have not received the attention they deserve in physics education journals; this work may be even the first claim for more attention to them. These functions of potential energy do not have the value of a physics fundamental principle, but their ubiquity represents for undergraduate and graduate students a systematic avenue of common conceptual attributes and a single formalism to learn. A formalism that can lead them to eventually exploit some of the different areas of research and applications emerged since 1990´s, or shall eventually emerge in the present decade, where these potentials are essential.

In Section 3 we presented the ladder of quantized eigenenergy levels of the phase quasi-particle $\varphi$ of a Josephson junction when trapped in an *anharmonic well* of its tilted washboard potential. Since each energy $E_i$ in that ladder corresponds to a different quantum state $|\varphi_i\rangle$. This provides opportunities for the teaching of introductory quantum mechanics, and for presenting applications and research, and even for understanding more advanced quantum phenomenae. In effect: (i) It is an excellent opportunity for introducing and teaching the Hilbert space of states of a quantum *nonlinear* oscillator; (ii) a single pair of these distinguishable quantum states allowed the conception and development of the *phase qubit*, and its application to build a quantum computer. These set of opportunities are significant arguments for having presented in this work the analytical derivation of the inclined washboard potential function of a Josephson junction, and showing its analogy with a classical driven-damped oscillator.

In Section 4 we presented two types of electro-optical washboard-like potentials generated with laser beams interacting with dielectric particles. Firstly, we introduced the Bessel laser beams and showed that they can be generated with a really low-cost optical setup by undergraduate students in a *medium* level, or in an *advanced* level physics teaching laboratory. We recognize that apart from the easy and low-cost experimental set-ups to generate these potentials, the theory and experimental set-ups to further investigate new optical and quantum phenomenae might not always lay within the scope of undergraduate students. But our second goal –as written in the Introduction– was just to offer beach-bridges for students to approach the applications of tilted washboard potentials research, or eventually to consider one of those applications as subjects of their eventual dissertations at college or graduate level. This goal we think has been achieved. Moreover, we have emphasized that many of the experimental set-ups –for the application of these washboards electro-optical potentials in present areas of research– are under straightforward and easy control of the experimentalist, an asset that favours the approach of physics and electronic majors to those applications. We have explained that the tilting of the transverse radial irradiance distribution of Bessel´s electro-optical potentials is due to the transverse Gaussian irradiance distribution of the laser beams used to irradiate the axicon (or the circular-slit aperture) when generating the Bessel beam, an explanation omitted in most publications on the subject.

Periodical electro-optical potentials generated with laser standing waves were also presented in Section 4; potentials being generated by overlapping two mutually coherent counter-propagating laser beams, with optical set-ups that again are within the reach of undergraduate physics and electronics students in their laboratories. The interaction of the resulting standing electric field intensity with sub-microscopic dielectric particles, *e.g. neutral* atoms, has led to the new key concept of *optical lattice*. An optical lattice trap atoms, at its narrow potential wells (micro traps) as if the lattice were a *crystal of light*. Their standing electric field intensity pattern, width and depth, are under complete control of the experimenter, and even 2-dim optical lattices can be generated. The latter lattices are usually 2-dim arrays of ~100000 or more potentials wells where neutral atoms, or even cells, may be trapped or submitted to controlled interactions by the experimentalists, *e.g.* motorized transportation along the lattice. Standing wave electro-optical potentials and its applications are right now (2023) subjects of MSc theses, and even of a remarkable Bachelor thesis [47]. Such is the case of clocks based on optical lattices that have accuracies far exceeding those of caesium atomic clocks. Optical lattices can be used for sensing our planet gravitational potential at high altitudes above the atmosphere with outstanding accuracy (~ 1cm altitude error). We hope to have provided instructors and undergraduates with a resource for enhancing their cognitive processes, and their eventual approach to the quantum and electro-optical advanced researches and technologies here mentioned.

**REFERENCES**

1. Tipler P A and Llewellyn R A 2008 *Modern Physics* 5th ed (Freeman, New York)
2. Griffiths D, 1994 *Introduction to Quantum Mechanics* (Prentice Hall, New Jersey)
3. Cohen-Tannoudji C, Diu B and Laloe F, 1977 *Quantum Mechanics*, Vol. 1 (Wiley, New York)
4. Tinkham M, 1996 *Introduction to Superconductivity*, 2nd ed., McGraw Hill (New York)
5. Tornes I E 2006 Topics in the Physics of Underdamped Josephson Systems, Ph D dissertation, Ohio State University
6. Tatarkova S A, Sibett W and Dholakia K 2003 Brownian Particle in an Optical Potential of the Washboard Type, *Phys. Rev. Lett* Vol. 3, 038101
7. Frisk A and Nori F 2019 *Quantum bits with Josephson junctions*, arXiv:1908.09558v1 [quant-ph]
8. De Luca R, Giordano A and D'Acunto I 2015 Mechanical analog of an over-damped Josephson junction, *Eur. J. Phys*. **36** 055042
9. Feynman R P, Leighton R B, and Sands M 1965 *The Feynman Lectures on Physics*, vol III ch 21 (Reading MA, Addison-Wesley)
10. McGloine D and Dholakia K 2005, Bessel Beams: Diffraction in a new light, *Contemp. Phys*. 46, 15.
11. Josephson B D 1964 *Rev. Mod. Phys.,***38,** 216
12. https//en.wikipedia.org/Josephson voltage−standard
13. Halliday D, Resnick R and Krane K S 2008 *Physics* 4th. ed Vol. II. ch.28
14. Jackson J D 2005 *Classical Electrodynamics* (Wiley, New York)
15. Ashkin A 1970 "Acceleration and trapping of particles by radiation pressure, *Phys. Rev. Lett*. **24**, 156–159 (1970)
16. Arfken G B and Weber J 1995 *Mathematical Methods for Physicists*, 4th ed., ch. 11 (New York, Academic Pr).
17. McQueen C A, Arlt J and Dholakia K 1999 An experiment to study a ''nondiffracting'' light beam *Am. J. Phys*., Vol. 67, No. 10
18. Bloch I 2005 Ultracold quantum gases in optical lattices, *Nature Physics,* **1,** 23-30, www.nature.com/naturephysics
19. Hand L N and Finch J D 1998 *Analytical Mechanics* (Cambridge UK, Cambridge University Press)
20. Bak C F and Pedersen N F 1973 Josephson junction analogue and quasi-particle particle, *App. Phys. Lett.*, **22**,149
21. Tuckerman D B and Phillips H 1980 JA-100 Josephson Junction Analog, (Stanford, Phillips-Gillete and Associates)
22. McConnell A, Idris S, Opatosky B and Amet F 2021 Phase locking and noise driven dynamics in a Josephson junction electronic analog, arXIv

23. Henry R W, Prober D E and A. Davidson A, 1981 Simple electronic analog of a Josephson junction, *Am. Jour. Phys,* 49
24. McCumber D E 1968 Effect of ac impedance on dc voltage-current characteristics of superconductor weak link junctions, *J. App. Phys* **19,** pp. 3113-3118
25. Stewart W C 1968 Current–voltage characteristics of Josephson junctions, *App. Phys. Lett*, **12**, 277-280
26. Feynman R P 1982 Simulating Physics with Computers (PDF). *International Journal of Theoretical Physics* **21** 467–488.
27. Clarke J and Wilhelm F K 2008 Superconducting quantum bits, *Nature* 453, 1031–1042. https://doi.org/10.1038/nature07128
28. Wendin G and Shumeiko S V 2005, Superconductors, Qubits and Computing, arXiv:cond-mat/0508729v1 [cond-mat.supr-con]
29. Ladera C L and Mata Toledo R 2021, Quantum Computation, *AccessScience* (New York, McGraw-Hill)
30. Nielsen M and Chuang I 2010 *Quantum Computation and Quantum Information* 10th ed.. doi:10.1017/CBO9780511976667
31. DiVincenzo D P 2000 The physical implementation of quantum computation, *Fortschritte der Physik* **48**, 771
32. McCleod J H 1954 The Axicon: a new type of optical element, *J. Opt. Soc. Am.* **44** , 592 - 597
33. Khonina S N, Kazanski N. L, Khorin P A and Butt M A, 2021Modern Types of Axicons: New Functions and Applications, *Sensors*, 21, 6690. https://doi.org/10.3390/s21196690
34. Hecht E, 1987 *Optics* 2$^{nd}$ed (Reading MA, Addison Wesley)
35. Wang Y, Yan S, Friberg A, Kuebel D and Visser T D 2017, Electromagnetic diffraction theory of refractive axicon lenses, *J .O.S.A* **A-34,** 120134
36. Durnin J 1987 Exact solutions for nondiffracting beams I, The scalar theory *J.O.S.A* **A- 4**, 651 – 654
37. ImageJ, National Institute of Health, USA
38. Milne G, Dholakia K, McGloin D, Volke-Sepulveda K and Zemánek P 2007 Transverse particle dynamics in a Bessel beam *Optics Express*, Vol. 13, 13973
39. Appleyard D C, Vandermeulen K Y, Hand L and Lang M J, 2007 Optical trapping for undergraduates *Am. J. Phys*. **7** 5-14
40. Guangham M, 2019 High-throughput volumetric imaging of neural dynamics in vivo, U. C. Berkeley, Ph. D. Thesis, https://escholarship.org/uc/item/21m0q3wp
41. Smith D E, Tans S J, Smith S B, Grimes S, Anderson D L, Bustamante C, 2001, The bacteriophage straight phi29 portal motor can package DNA against a large internal force., *Nature*. **413** (6857): 748–52
42. Galvez E J and Zhelev N 2007 Orbital Angular Momentum of Light in Optics Instruction, OSA Technical Digest Series, https://doi.org/10.1364/ETOP.2007.ESB3
43. French A P 1971 *Vibrations and Waves* (Nelson, London)
44. Crawford F S 1966 *Waves,* Vol. 4, Berkeley Physics Course,Vol.3,(McGraw Hill, New York)

**45**. Grimm G and Muller M and Ovchinnikov Y 1999 Optical traps for neutral atoms, arXiv:physics/9902072v1
**46**. Uerlings P M 2019, Phase Stability of an optical superlattice setup for ultracold dysprosium atoms, *B. Sc. Thesis,* (Germany, University of Stuttgart)
**47.** Wili S, Tilman Esslinger T, and Viebahn K 2023 An accordion superlattice for controlling atom separation in optical potentials arXiv:2301.04144v1 [cond-mat.quant-gas]
**48.** Katori H, Takamoto M, Pal'chikov V G, Ovchinnikov V D 2003, Ultrastable Optical Clock with Neutral Atoms in an Engineered Light Shift Trap, *Physical Review Letters*. **91** (17): 173005. arXiv:physics/0309043
**49**. Kolkowitz S, Pikovski I, Langellier N, Lukin M D, Walsworth R.L, Ye J 2016, Gravitational wave detection with optical lattice atomic clocks arXiv:1606.01859v3 [physics.atom-ph]